\documentclass[oldversion]{aa}  
\usepackage{graphicx}
\usepackage{txfonts}
\usepackage[]{natbib}
\begin{document}
   \title{Optical colours of AGN in the Extended \emph{Chandra} Deep Field South:
          Obscured black holes in early type galaxies}

   \authorrunning{Rovilos \& Georgantopoulos}
   \titlerunning{Optical colours of AGN in the ECDFS}

   \author{E. Rovilos,\inst{1,2} 
           \and
           I. Georgantopoulos,\inst{1}
              }

   \offprints{E. Rovilos\\ \email{erovilos@astro.noa.gr}}

   \institute{Institute for Astronomy and Astrophysics, National
             Observatory of Athens, I. Metaxa \& V. Pavlou str, Palaia
             Penteli, 15236, Athens, Greece
         \and
             Astronomical Laboratory, Department of Physics, University
             of Patras, 26500, Rio-Patras, Greece
             }

   \date{Received date; accepted date}

    \abstract{We investigate the optical colours of X-ray sources
from the Extended \emph{Chandra} Deep Field South (ECDFS) using photometry
from the COMBO-17 survey, aiming to explore AGN - galaxy feedback models.
The X-ray sources populate both the ``blue'' and the ``red sequence'' on the
colour-magnitude diagram. However, sources in
the ``red sequence'' appear systematically more obscured. \emph{HST} imaging
from the GEMS survey demonstrates that the nucleus does not affect
significantly the observed colours, and therefore red sources are early-type
systems. In the context of AGN feedback models, this means that there is
still remaining material after the initial ``blowout''. We argue that this
material could not be only left-over from the original merger, but a secondary
cold gas supplier (such as minor interactions or self-gravitational
instabilities) must also assist.
}


   \maketitle
%

\section{Introduction}

The interplay between star formation and AGN activity in galaxies is one of
their most striking properties. Both phenomena are often linked with merger
events \citep[e.g.][]{Larson1978,Stockton1982} and this was for years
considered their only connection. The mass of the black hole
is found to be tightly correlated with properties of the host galaxy, such as
the bulge luminosity \citep{Magorrian1998}, its mass \citep{Merritt2001} and
its velocity dispersion
\citep[$M_{\star}-\sigma$ relation;][]{Ferrarese2000,Gebhardt2000} suggesting
that there is a close connection of the central engine with its host galaxy.
These observational trends have been taken into account in early,
semi-analytical models of AGN - host galaxy co-evolution
\citep[e.g.][]{Kauffmann2000}, which managed to predict many of their
observational characteristics, such as the quasar luminosity function and its
evolution.

Modern evolutionary models of AGN track the evolution of major merger events,
which are common in redshifts $z\gtrsim 2.5$ \citep{Conselice2003}. They take
into account feedback from the starburst itself through supernova explosions
and the central super-massive black hole \citep{Granato2004,Monaco2004}.
Feedback can heat the cold gas supply which feeds both the starburst and the
AGN, thus regulating the activity in galactic centres. AGN and QSO evolution
models accounting for feedback
\citep[e.g.][]{Hopkins2005a,Hopkins2005b,Hopkins2006} predict that the AGN is
obscured for most of its lifetime \citep[see also][]{Page2004} and is directly
observable only at the later stages, when the majority of cold gas has been
swept away and before it stops accreting because of fuel shortage. Such a
scenario supports previous claims giving similar predictions
\citep{Sanders1988}. At the last stages of its evolution, the AGN can maintain
low level activity by accreting hot surrounding gas in a so called `radio mode'
\citep{Croton2006} or from fresh gas supply as a result of interactions with
other systems \citep*{Cavaliere2000,Mouri2004,Menci2004,Vittorini2005}.

A useful tool for testing the implications of such models is the inspection of
the optical colours of the galaxies. A bimodal distribution has been observed
in the colour-magnitude diagram \citep{Baldry2004,Bell2004}, which is explained
with star formation \citep[e.g.][]{Menci2005}. The red cloud is populated
by old systems that are passively evolving, whereas blue galaxies owe their
colour to active star formation. According to popular models of AGN evolution,
galaxies start their lives in the blue cloud and they migrate to the red
sequence when star formation is quenched and the AGN can be directly viewed.
\citet{Nandra2007} used the colour distribution of a number of X-ray selected
AGN in the AEGIS survey \citep{Davis2007} to show that they are preferably
located in the red sequence and the ``valley'', between the red sequence
and the blue cloud. According to these authors, this reflects the emergence
of the AGN only at the latest stages of quasar evolution. However, there
still remain a few issues that need to be addressed. The obscuration status
of X-ray selected AGN poses a problem when interpreting it with standard models
and the contribution of the nucleus to the colour of the galaxy is still
unclear.

In this paper, we use a large number of X-ray sources from the ECDFS survey
\citep{Lehmer2005} to revisit this issue. We also use \emph{HST} observations
\citep{Rix2004} to evaluate the contamination of the central source to the
optical colours. We adopt $\mathrm{H_{0}=72\,km\,s^{-1}\,Mpc^{-1}}$,
$\Omega_{\mathrm{M}}=0.3$ and $\Omega_{\Lambda}=0.7$.

\section{Data}

\begin{figure}[h]
   \centering
   \includegraphics[scale=0.45]{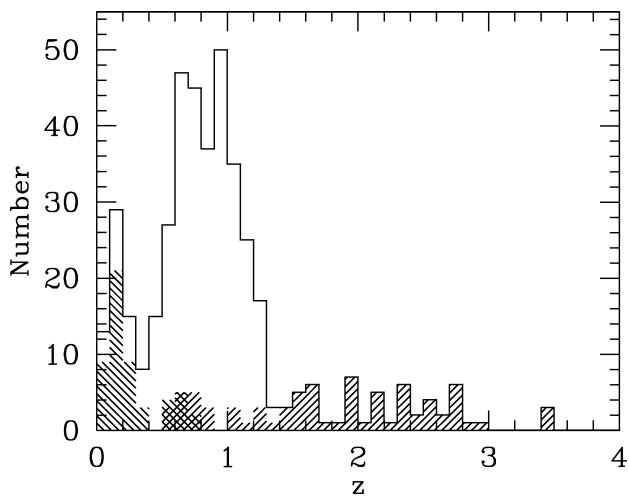}
   \caption{Redshift distribution of the various source types. The overall
            histogram refers to all sources, whereas the shaded refer to normal
            galaxies (lower redshifts) and optical QSOs (higher redshifts).}
   \label{redshifts}
   \end{figure}

We select our X-ray sources using the public catalogue of the Extended
\emph{Chandra} Deep Field South \citep[ECDFS;][]{Lehmer2005}, which reaches
depths of $1.1\times10^{-16}$\,erg\,cm$^{-2}$\,s$^{-1}$ and
$6.7\times10^{-16}$\,erg\,cm$^{-2}$\,s$^{-1}$ in the $(0.5-2.0)$\,keV and
$(2-8)$\,keV bands respectively. We combine these data with the optical
catalogue of the ECDFS from the COMBO-17 survey \citep{Wolf2004}, which
provides photometric redshifts, as well as rest-frame optical colours for
most of the sources. Within a search radius of 3\,arcsec, we find optical
counterparts for 578 of the 762 ECDF-S sources, 421 of which with calculated
photometric redshifts. The redshift distribution of those sources is shown
in Fig. \ref{redshifts} (main histogram). Among the X-ray sources in the
ECDFS there are normal galaxies, which do not host an AGN. We use X-ray
criteria ($L_{\rm x}<10^{42}$\,erg\,s$^{-1}$; $HR<-0.2$; see \citealt{Bauer2004})
to identify them and remove them from our AGN sample. As can be seen in Fig.
\ref{redshifts} (shaded histogram) they are confined in redshifts $z<0.5$.

The COMBO-17 catalogue provides rest-frame optical colours for all sources
whose SED is fitted with a ``galaxy'' template but not for optical QSOs where
the light of the AGN dominates the broad-band spectrum. In those cases we
calculated the rest-frame colours assuming a power-law spectrum with slope
$\alpha=-1$ and corrected the results by 0.5\,mag. This method introduces an
r.m.s. scatter of 0.24\,mag according to \citet{Wolf2003}, which is reasonable
for the scope of this study, as optical QSOs are a small fraction, 21\% in all
redshifts and 7\% in redshifts $0.6<z<1.2$. Moreover, they occupy a distinct
region of the colour-magnitude diagram and their optical colours are dominated
by the nuclear regions (see next paragraphs), therefore they are not considered
for statistical evaluations.

In the following discussion, we will focus on the redshift range of
$0.6<z<1.2$, unless otherwise stated to encompass the peak of the redshift
distribution of AGN participating in the formation of the X-ray Background
\citep{Barger2005}. This redshift selection includes 63\% of sources with a
super-massive black hole (moderate AGNs and QSOs) and 74\% of moderate AGNs
(fitted with a ``galaxy'' template - non QSOs). As can be seen in Fig.
\ref{redshifts}, at $z<0.6$ there is the bulk of normal galaxies
\citep[see also][for the central region]{Zheng2004} and at $z>1.2$ there are
many optical QSOs \citep[see also][]{Wolf2004}.

\section{Results}

   \begin{figure*}
   \centering
   \includegraphics[scale=0.65]{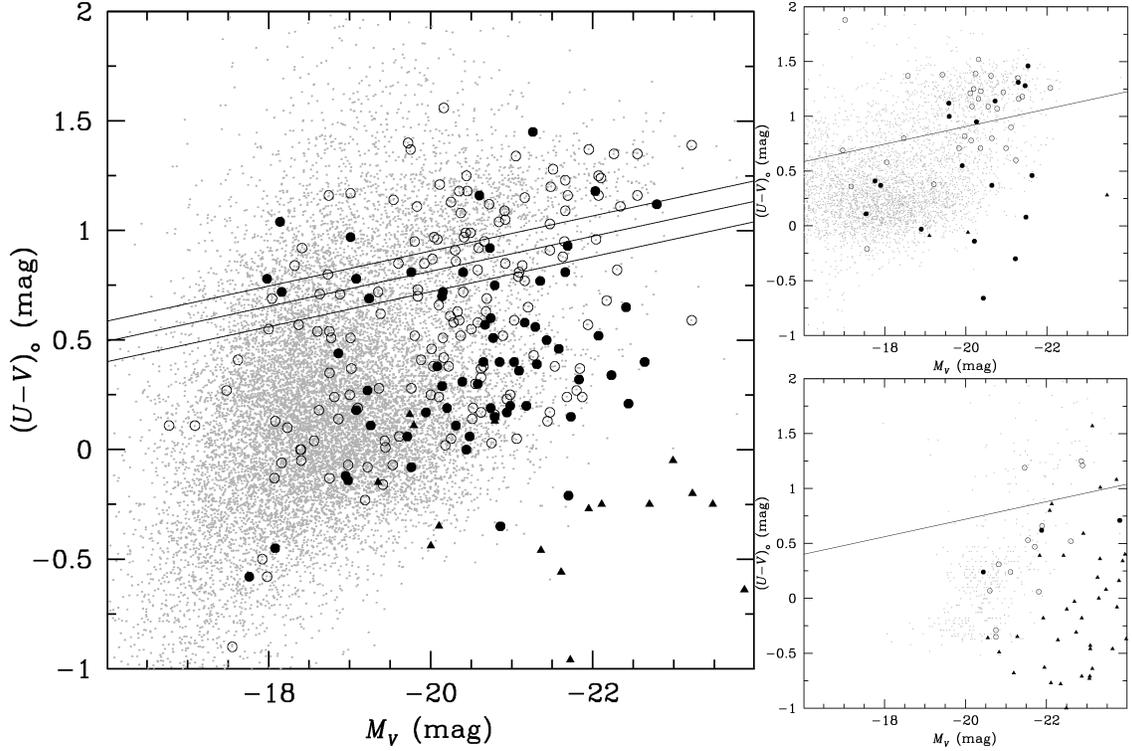}
   \caption{Colour-magnitude diagram of X-ray source of the Extended
            \emph{Chandra}
            Deep Field South with $0.6<z<1.2$ (main panel). Hard sources
            $(HR>-0.2)$ are shown as open circles and soft $(HR<-0.2)$ as
            filled circles. COMBO-17 sources in the same redshift range are
            shown in gray dots. Sources fitted with a QSO template in COMBO-17
            are shown in filled triangles. The lines represent the limits of
            the red sequence for three redshifts, 0.6, 0.9, and 1.2, according
            to \citet{Bell2004}. The left upper and lower panels show the
            redshift ranges $z<0.6$ and $z>1.2$ using the same symbols.}
   \label{CMD}
   \end{figure*}

The colour-magnitude diagram is often used to examine the evolution of
different kinds of galaxies \citep[see][]{Baldry2004,Bell2004}.
We plot the Johnson $(U-V)$ optical rest-frame colour versus the $V$
rest-frame magnitude (Vega magnitudes) in Fig. \ref{CMD}. The main (left)
diagram refers to sources in the main redshift range ($0.6<z<1.2$), whereas
the diagrams on the right refer to sources with $z<0.6$ (upper) and $z>1.2$
(lower). We use different symbols for plotting obscured and unobscured sources
according to their hardness ratios. Open circles mark sources with
$HR>-0.2$ (corresponding to $N_{\rm H}\simeq 2.4\times 10^{22}\,{\rm cm}^{-2}$
for $z=1$ and intrinsic $\Gamma=1.9$) and filled circles mark sources with
$HR<-0.2$, whereas sources with QSO templates are marked with a filled
triangle (all have $HR<-0.2$) and occupy a distinct region in the CMD.

\begin{figure}[h]
   \centering
   \includegraphics[scale=0.45]{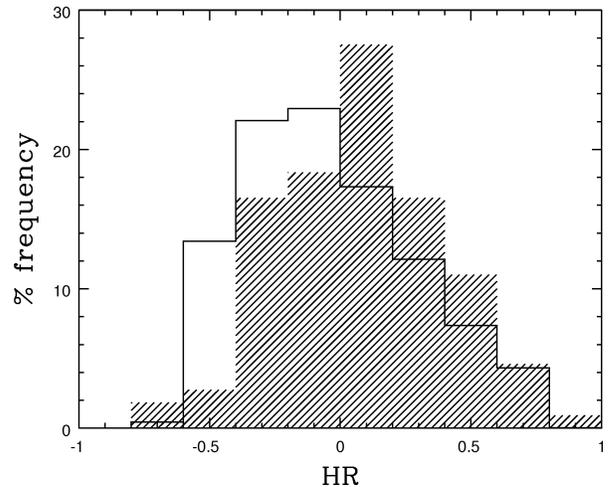}
   \caption{Hardness ratio histograms of the red (shaded) and the remaining
            sources.}
   \label{HR}
   \end{figure}

In Fig. \ref{CMD} we can see that sources in the main redshift range and in the
red sequence appear harder than those in the blue cloud and the region in
between. Only 9 (14\%) of the unobscured sources ($HR<-0.2$) have red colours
\citep[$U-V>1.15-0.31z-0.08(V-5\log h+20)$;][]{Bell2004} while 28\% of all
sources are red. The hardness ratio histograms for blue (shaded) and red
sources are shown in Fig. \ref{HR}, where although both distributions have
large scatters \citep[see also][]{Akylas2006}, it is clear that red sources are
generally harder. To statistically evaluate this trend, we use the Kolmogorov
- Smirnov test and find that within 99.9\% the blue and red sources are drawn
from different populations in terms of hardness ratio. However,
there is a number of sources having upper or lower limits in their hardness
ratios, which might affect this result. To overcome this we created a hard
and a soft sub-sample of our sources (with hardness ratio lower and upper
limits respectively) and performed the Gehan's test to each sub-sample
separately to check if the blue and red sources are drawn from different
populations in terms of $HR$. In both cases the probability is higher than
98.7\%. For the sources outside the main redshift range, the statistical
significance of this trend is much lower, the null hypothesis is 34.6\%
and 10.6\% probable for low and high redshift sources respectively. The
fact however remains that we can see a significant number of obscured sources
in the red cloud.

The ratio of X-ray obscured to unobscured sources is related to the X-ray
luminosity \citep{Ueda2003,Akylas2006}, as galaxies with lower X-ray luminosity
tend to be more obscured. The association of red systems with enhanced X-ray
obscuration could affect the X-ray luminosity, if the red sources
were less X-ray luminous. To estimate the effect of this we calculate the X-ray
luminosities in the $0.5-8$\,keV band for the AGN of Fig. \ref{CMD} (left).
This X-ray band is affected by absorption, and we have applied a correction
based on the hardness ratio of each source. We have assumed a power-law X-ray
spectrum with an intrinsic $\Gamma=1.9$ \citep{Nandra1994}, obscured by an
optimum hydrogen column density ($N_{\rm H}$) to reproduce the observed hardness
ratio.

\begin{figure}
   \centering
   \includegraphics[scale=0.5]{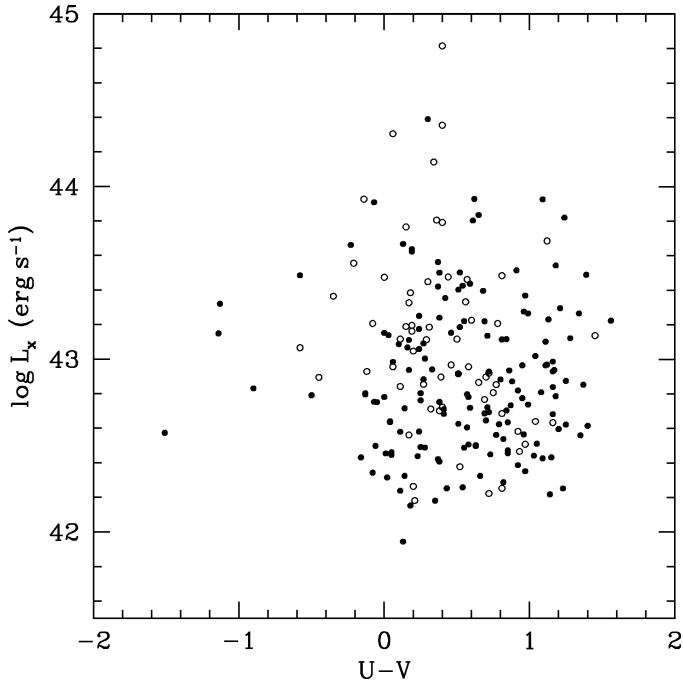}
   \caption{X-ray luminosity against optical colour for AGN with redshifts
            $0.6<z<1.2$. Hard (soft) sources are shown as filled (open)
            circles.}
   \label{Lx}
   \end{figure}

In Fig. \ref{Lx} we plot the intrinsic X-ray luminosity against the $U-V$
colour. The soft X-ray sources (with $HR<-0.2$) are more luminous than the hard
(with $HR>-0.2$), at the 99.7\% significance level.
However, we do not observe any correlation between the X-ray luminosity and the
optical colour. The average luminosity of the AGN in the red cloud is
marginally lower than this of the blue, but the probability that those two
populations are drawn from the same parent distribution is 54.5\%, which does
not allow any correlation. Therefore, we conclude that the obscuration detected
toward red galaxies is not a luminosity effect.

An issue which still remains unresolved is the impact the nucleus has in the
optical colours. High resolution optical images have been taken with the
\emph{HST} by \citet{Rix2004} in the GEMS survey (Galaxy Evolution from
Morphology and SEDs). This survey covers the outskirts of the ECDFS in $V$
(F606W) and $z$ (F850LP), while the inner region is observed with the
\emph{HST} as part of the GOODS survey. These data have been re-reduced by the
GEMS team to obtain a smooth field. Optical inspection of the $V$ and $z$
images reveals that red sources are mostly associated with early type galaxies,
in agreement with \citet{Bell2004b}, while the nuclear region affects the
sources with the bluest optical colours.

   \begin{figure}[h]
   \centering
   \includegraphics[scale=0.45]{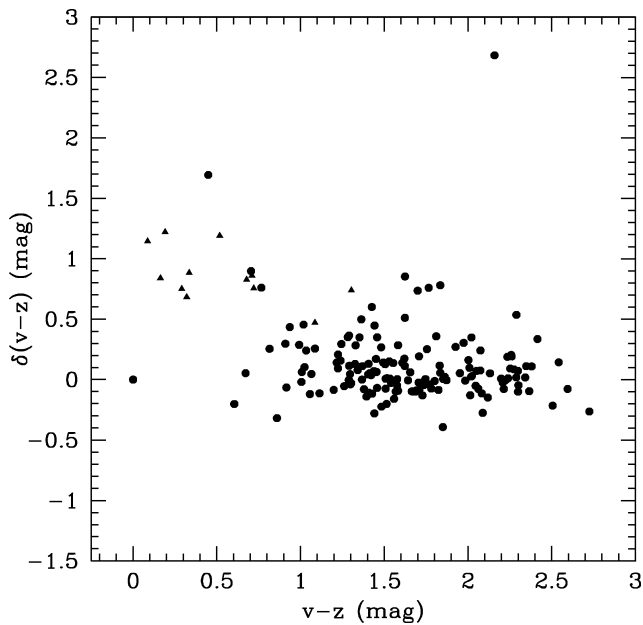}
   \caption{Difference in observer's frame $V-z$ colour if we exclude the
            central region of the source. Optical QSOs are plotted in
            triangles}
   \label{colours}
   \end{figure}

We used the $V$ and $z$ GEMS images to estimate the contribution of the
nucleus to the colour of the system. We conducted photometry to the optical
sources related to X-ray sources and repeated this excluding a central
region of radius 0.3\,arcsec ($\simeq 2.3$\,kpc for $z=1$) and compared the
results. The change in the
$V-z$ (observer's frame) colour with respect to it is shown in Fig.
\ref{colours}. We can see that although there is some scatter in the colour
difference around its mean value, this remains
close to zero. Only 11 sources (7.1\%) have $\delta(V-z)>0.5$
and we do not detect any sources with red nuclei which affect the
overall colour. Therefore
we can safely regard optical rest-frame $U-V$ colours (roughly
corresponding to observer's frame $V-z$ for $z\approx1$) as a property of the
host galaxy and not the AGN. An exception are optical QSOs
which as can be seen in Fig. \ref{colours} owe their colour to the blue
nucleus.

Since the nuclear light has little impact on the overall colour of the system,
this can be attributed to either  an old stellar population or a dusty
starburst. \citet{Roche2003} find that many red galaxies in the CDFS/GOODS
fields have large amounts of dust which define their optical colours. Dust
lanes away from the AGN \citep[e.g.][]{Radomski2003} could redden the system
and still be far from the nucleus, so that the nuclear light would not dominate
the overall colour. However, optical inspection of their GEMS morphologies
reveal early type structures, according to the findings of \citet{Bell2004b}.
We therefore assume that dust has a minimal contribution to the optical colours
of the majority of sources in the red cloud.

   \begin{figure}[h]
   \centering
   \includegraphics[scale=0.45]{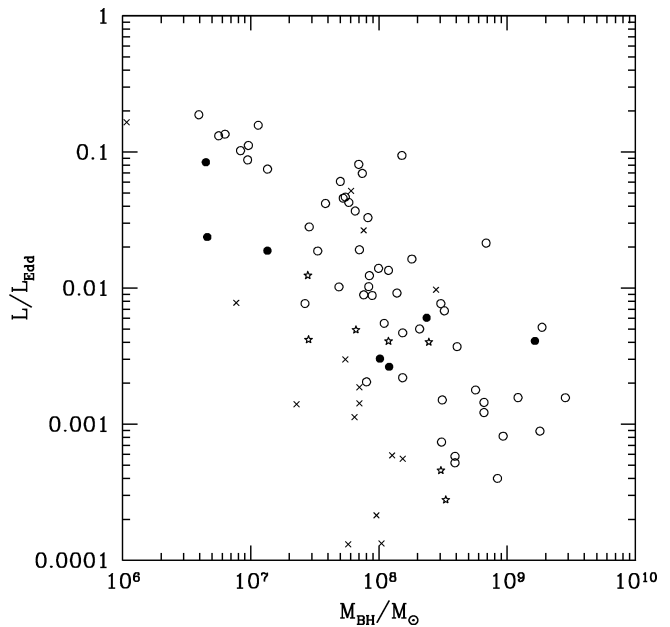}
   \caption{Eddington rate versus central black hole mass for the AGN in the
            red cloud. Open (filled) circles represent obscured (unobscured)
            sources in the main redshift range ($0.6<z<1.2$), whereas obscured
            and unobscured sources with lower redshifts $z<0.6$ are plotted
            with crosses and stars respectively.}
   \label{edd}
   \end{figure}

So far we have seen that AGN having red colours have late-type morphologies
(are thus evolved) and at the same time show evidence of X-ray obscuration.
According to galaxy-AGN evolution models invoking AGN (and/or starburst)
feedback \citep[e.g.][]{Hopkins2006}, AGN should shine unobscured at the latest
stages of their evolution, as a result of the ``blowout'' of the obscuring
material. However, AGN feedback has a lower efficiency at lower black hole
masses \citep[e.g.][]{Fabian1999,Granato2004,Shankar2006}, so we do expect some
residual
obscuring material when the black hole mass is relatively low. To test this
assumption, we need to estimate these masses. The galaxies in the red cloud in
our AGN sample have elliptical-type morphologies, so their optical luminosities
can reveal the masses of their nuclear sources \citep{Magorrian1998}. We use
the $V$ band luminosity and calculate:
$\log(M/M_{\odot})=-0.55(M_{\rm V}+22)+8.78$, according to
\citet{Lauer2007}. The X-ray luminosity on the other hand is tightly related
to the total bolometric luminosity, depending on the accretion rate. We use the
luminosity-dependent bolometric correction from \citet{Hopkins2007}, using the
hard ($2-10\,keV$) X-ray luminosity (which is less affected by obscuration) to
derive the bolometric luminosity of the AGN:
$L_{\rm bol}/L_{\rm 2-10\,keV}=10.83(L_{\rm bol}/10^{10}L_{\odot})^{0.28}+6.08(L_{\rm bol}/10^{10}L_{\odot})^{-0.02}$. We can then calculate the Eddington rate at which
mass is accreted to the central black hole as:
$L_{\rm bol}/L_{\rm Edd}\sim 3\times 10^{5}L_{\rm bol}/10^{-0.55(M_{\rm V}+22)})$,
where $L_{\rm x}$ is given in erg\,s$^{-1}$.

In Fig. \ref{edd} we plot the Eddington rate against the mass of the central
black hole for red sources. Obscured sources (having $HR>-0.2$) within the main
redshift range ($0.6<z<1.2$) are plotted with open circles and unobscured
with filled circles. Sources with $z<0.6$ are plotted with crosses and stars if
they are obscured and unobscured respectively. We note that the masses of the
black holes of the red sources span across three orders of magnitude,
reaching few times ${\rm 10^9M_\odot}$, characteristic of AGNs
\citep[see][]{Shankar2004}, so they are not considered the low end of the black
hole mass function. Moreover, the obscured sources do not have lower masses
compared with those which show no signs of X-ray obscuration, suggesting that
the obscured red sources
are not a result of lower black hole masses and a limited ``blowout''
efficiency. In Fig. \ref{edd} we can also see that the red AGN are accreting
at sub-Eddington rates (generally $\lesssim 0.1$), while there is an obvious decrease in the Eddington
rate with increasing black hole mass. This is a hint that the accretion rate
is independent of the black hole mass, so it is probably regulated by the gas
supply and not the size of the central black hole for galaxies in the red
sequence. Lower redshift sources also have lower Eddington rates than
those in the main redshift range, for their respective masses. For bluer
sources we cannot estimate the black hole masses and Eddington rates because
their morphologies are more complicated and a selection of the bulge part of
the optical emission would be needed.

\section{Discussion}

The colour magnitude diagram (CMD) provides a useful tool to explore the
evolutionary status of moderate luminosity AGN. X-ray selected sources tend to
be optically more luminous than COMBO-17 sources (Fig. \ref{CMD}) and occupy
the red end of the blue cloud, the red cloud and the region in between.
\citet{Nandra2007} observed a similar distribution of X-ray sources in the
AEGIS survey and interpreted it using models of QSO evolution based on AGN
feedback \citep[e.g.][]{Hopkins2006}.
These models describe the migration of galaxies from the blue cloud to the red
sequence when AGN activity becomes powerful enough to quench star formation
sweeping away the obscuring material. However, the existence of a number of
obscured sources in the red cloud in the \citet{Nandra2007} sample could not be
straightforwardly explained with AGN feedback models, which predict that the
AGN is unobscured at the latter stages of its evolution linking the obscuring
with the star-forming gas. We extend this result by showing that red galaxies
have early-type morphologies (they are not dust reddened) and are
\emph{preferentially} more obscured.

One important question that needs to be raised before associating early-type
galaxies with obscured nuclei, is whether there are unobscured early-type
galaxies in the blue cloud, just because their colours are contaminated by
their nuclear light. There are some objects in Fig. \ref{colours} deviating
from the average `galaxy' colours, and the most extreme cases
(those with $\delta(V-z)>0.7$) are all not obscured members of the blue cloud
according to the \citet{Bell2004} criterion and have morphologies resembling
early-type galaxies. However, their number is small (7 sources out of 165 with
GEMS counterparts) and they do not affect our result in great extent.

Since a red colour implies an old stellar population, there has to exist a
mechanism that obscures the X-rays after the initial ``blowout'' of foreground
cold gas by AGN feedback. The most obvious scenario has to do with the
efficiency of the feedback. It is possible that lower mass AGN are not powerful
enough to completely disrupt the surrounding gas, so residual material could
both obscure and feed the AGN. The redshift range probed in this paper
($0.6<z<1.2$) is lower than this where massive galaxies are formed,
and at lower redshifts the dark matter potential wells which reach
virialization are shallower, as a result of the decrease in virialization
time with the mass of the protogalaxy \citep{Granato2004}. However, the black
hole masses we measure are not at the low end of the black hole mass function
and obscuration seems not to be related with the black hole mass (see Fig.
\ref{edd}). Moreover, The X-ray luminosity of red sources is not less than this
of blue sources, which means that the accretion rates of evolved systems are
not lower. Both these make the ``blowout efficiency'' scenario less likely,
however without ruling out that at least part of the obscuring gas is left-over
from the initial blowout, especially for lower mass black holes.

If the original merger-driven material is disrupted as a result of AGN
feedback, there has to be a secondary gas supply to both sustain AGN activity
and obscure the AGN at later epochs. \citet{Croton2006} propose a `radio mode'
for the latest stages of quasar evolution, where AGN activity is preserved by
the accretion of hot gas from the newly-formed galactic halo. The destruction
of the star-forming cool gas \citep[e.g.][]{King2005} by the AGN causes the
migration to the red sequence, while the hot halo continues to feed the AGN.
This could explain the obscured red sources (although not the correlation
between obscured AGN and red sources) if we assume that the hot halo can
provide large enough column densities to obscure the X-rays. However, in such a
scenario, galaxies in the red sequence would again be less luminous in X-rays,
as a result of the lower accretion rates \citep{Croton2006}. The X-ray
luminosities of the sources in the red cloud are not smaller than these of the
blue cloud (see Fig. \ref{Lx}), disfavoring such a hypothesis and requiring a
secondary cold gas supplier to re-fuel the system.

This secondary cold gas supply which revives the AGN and obscures the system
could be a result of minor interactions with nearby galaxies or even self
gravitational instabilities. Galaxy evolution models
consider major mergers as the mechanism which transfers large amounts of cold
gas into the central regions of galaxies, which is accreted into the black hole
enlarging it, or even sometimes generating it. At the same time this cold gas
triggers star formation episodes further away from the black hole. In cases
of minor galaxy interactions, which are more common in redshifts $z<2$, the
gas supply is limited, and could not be enough to sustain star formation,
while being enough to feed the AGN, as it needs a lower rate of gas supply to
function \citep{Mouri2004}. This cold gas could also provide the column density
needed for obscuration. These AGN-dominant Seyferts have gone through the
starburst-dominated phase, where the black hole experienced its main growth
phase and are now passively evolving, having a more early-type morphology,
like the red sources in our sample. Moreover, as their central black holes
have already grown, they are accreting material at sub-Eddington rates, while
the accretion is regulated by the gas supply and not the black hole mass. The
decline of the Eddington rate (being significantly lower than one) with black
hole mass seen in Fig. \ref{edd} supports this. Also, as seen in Fig.
\ref{edd}, red systems at lower redshifts ($z<0.6$) have lower Eddington
rates for their respective black-hole masses, which is consistent with the
decline of the galaxy interaction and merger rate with decreasing redshift
\citep[e.g.][]{LeFevre2000}.

The association of red galaxies with absorbed AGN has been witnessed earlier.
\citet{Silverman2005} associates X-ray absorbed AGN hosts with red galaxies
and finds evidence that the red colours reflect the presence of an early-type
galaxy and not a reddened AGN \citep*[see also][]{Georgakakis2006}. Here, we
further argue that the AGN does not influence the optical colours by directly
measuring its contribution with \emph{HST} imaging.
Our early-type absorbed AGN population may bear some similarities with the
Extremely Red Objects (ERO) population. \citet{Brusa2005} find that
almost all of ERO X-ray sources with redshift information from the literature
are obscured with $N_{\rm H}>10^{22}\,{\rm cm}^{-2}$
\citep[see also][]{Severgnini2005}.
While our sources are not typically EROs\footnote{The red sources in our
sample with available $R$ and $K$ measurements \citep{Szokoly2004} have
$R-K\simeq 4.5$, while EROs have $R-K>5$. Moreover, our red sources have
X-ray to optical flux ratios typical of an AGN
\citep[$-1<\log\frac{f_{\rm x}}{f_{R}}<1$;][]{Lehmann2001},
while EROs often have $\log\frac{f_{\rm x}}{f_{R}}>1$ \citep{Mignoli2004}},
they share some observational characteristics (colours redder than the
respective median, X-ray absorption) and could be their analogous sources in
more moderate redshifts
\citep[EROs usually have $z\gtrsim 1$;][]{Georgakakis2005}, at least of those
not being dust reddened. Red evolved galaxies are therefore rather common,
and theoretical models of galaxy and AGN evolution should be fine-tuned to
predict such a behavior in evolved systems.

\section{Conclusions}

In this paper we used the rest-frame $U-V$ colours from the COMBO-17 survey
to investigate the optical properties of X-ray selected ECDFS sources. We used
\emph{HST} imaging from the GEMS survey to examine the contribution of the
nucleus, which we found minimal for moderate optical luminosity sources (not
QSOs). We found that red sources tend to be more
obscured in X-rays, while they are linked with optical early-type systems.
This result enhances previous results finding X-ray obscuration in the red
cloud \citep{Nandra2007}. Merger driven AGN-galaxy co-evolution models do
predict obscuration in evolved systems after the effect of AGN feedback, but
require lower black hole masses and accretion rates, something which is not
supported by our data. However, fresh cold gas could re-fuel the system (e.g.
through interactions with a nearby galaxy or even gravitational instabilities
within the system) and in conjunction with residual cold gas from an incomplete
blowout it could revive the AGN and obscure the system. In this case the
accretion rate is regulated by the gas supply and the system is accreting at
low Eddington rates, as demonstrated by our results. There
a significant number of sources which are old and obscured, and theoretical
models should consider this observational trend.

\begin{acknowledgements}
ER wishes to thank the European Social Fund (ESF), Operational Program for
Educational and Vocational Training II (EPEAEK II), and particularly
the Program PYTHAGORAS II, for funding part of the above work. We thank A.
Georgakakis for useful discussions and the use of his software. We also thank
the anonymous referee for comments that improved the original manuscript.
\end{acknowledgements}

\end{document}